# SURVEY OF TRUST MODELS IN DIFFERENT NETWORK DOMAINS

Mohammad Momani [1], Subhash Challa [2]

[1]Faculty of Engineering and Information Technology, UTS, Sydney, *Australia*
mohammad.momani@uts.edu.au
[2]NICTA, VRL, University of Melbourne, Australia
*Subhash.Challa@nicta.com.au*

## ABSTRACT

*This paper introduces the security and trust concepts in wireless sensor networks and explains the difference between them, stating that even though both terms are used interchangeably when defining a secure system, they are not the same. The difference between reputation and trust is also explained, highlighting that reputation partially affects trust. A survey of trust and reputation systems in various domains is conducted, with more details given to models in ad-hoc and sensor networks as they are closely related to each other and to our research interests. The methodologies used to model trust and their references are presented. The factors affecting trust updating are summarised and some examples of the systems in which these factors have been implemented are given. The survey states that, even though researchers have started to explore the issue of trust in wireless sensor networks, they are still examining the trust associated with routing messages between nodes (binary events). However, wireless sensor networks are mainly deployed to monitor events and report data, both continuous and discrete. This leads to the development of new trust models addressing the continuous data issue and also to combine the data trust and the communication trust to infer the total trust.*

## 1. INTRODUCTION

Wireless sensor networks (WSNs) in recent years, have shown an unprecedented ability to observe and manipulate the physical world, however, as with almost every technology, the benefits of WSNs are accompanied by a significant risk factors and potential for abuse. So, someone might ask, how can a user trust the information provided by the sensor network?

Sensor nodes are small in size and able to sense events, process data, and communicate with each other to transfer information to the interested users. Figure 1 below, shows a typical sensor node (mote) developed by researchers at UC Berkeley called Mica2 as presented in [1] Typically, a sensor node consists of four sub-systems [2, 3].

- Computing sub-system (processor and memory): responsible for the control of the sensors and the execution of communication protocols.
- Communication sub-system (transceiver): used to communicate with neighbouring nodes and the outside world.
- Sensing sub-system (sensor): link the node to the outside world.
- Power supply sub-system (battery): supplies power to the node.

WSNs are a collection of self-organised sensor nodes that form a temporary network. Neither pre-defined network infrastructure nor centralised network administration exists. Wireless nodes communicate with each other via radio links and since they have a limited transmission range, nodes wishing to communicate with other nodes employ a multi-hop strategy for communicating and each node simultaneously acts as a router and as a host.

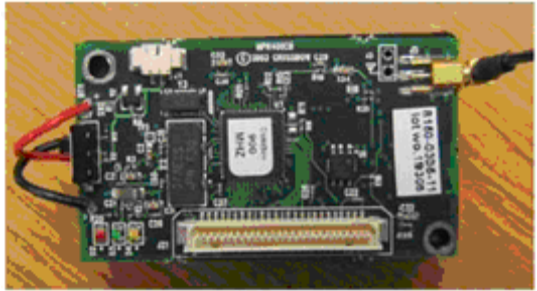

Figure 1. Sensor node example (mica2)

It should be noted that bandwidth available between communicating wireless nodes is restricted. This is because wireless networks have a significantly lower data transmission capacity compared to fixed-line data networks. Furthermore, wireless nodes only have a limited power supply available, as power supplied by batteries is easily exhausted. Lastly, wireless nodes may join or leave a network at any given time and frequently change their location in a network; this results in a highly dynamic network topology.

WSNs consist of sensor nodes with limited computation and communication capabilities deployed in large numbers, that is, tens of thousands as opposed to tens or hundreds of nodes. They present the same challenges that any other Mobile ad-hoc network (MANET) presents (absence of infrastructure, mobility, lack of guaranteed connectivity), but the computation constraint makes the design of solutions even harder. Also, WSNs have an additional function to the traditional functions of MANETs, which concerns monitoring events, collect and process data and transmit sensed information to interested users. This observed difference is the foundation of this new research to model trust in WSNs.

WSNs technology is a relatively new and emerging concept and has received increasing attention due to the advancement in wireless communications in the last few years. In addition, the need to have very tiny and cheap nodes being deployed in large numbers and in difficult environments such as military zones led to increased focus by researchers on WSNs. While wireless communication is already used in all sectors of daily life, WSNs have yet to step beyond the experimental stage. There is a strong interest in the deployment of WSNs in many applications and the research effort is significant.

Due to impressive technological innovations in electronics and communications, small low-cost sensor nodes are available, which can collect and relay environmental data [4, 5]. These nodes have sensing, computing and short-range communication abilities and can be deployed in many environments. Such deployment can be in controlled environments, such as sensing the atmosphere in buildings and factories, where the mobility of the nodes is of interest. Or they can be spread in hazardous and hostile environments and left unattended. Originally motivated by surveillance in battlefields for the military, interest in WSNs spread over a wide range of applications, from scientific exploration and monitoring, for example, the deployment of a WSN on an active volcano [6, 7], to monitoring the microclimate throughout the volume of redwood trees [8], to building and bridge monitoring [9, 10], to health-care monitoring [11] and a number of other applications such as [3, 4, 12, 13] such as:

- **Industrial Control and Monitoring**: control of commercial lighting, detect the presence of dangerous materials, control of the heating, ventilating, and air conditioning of buildings.

- **Home Automation**: design of the universal remote control, that can control not only the television, DVD player, stereo, and other home electronic equipment, but the lights, curtains, and locks. Personal computer peripherals control, such as wireless keyboards and mice. Remote keyless entry feature found on many automobiles.

- **Security and Military Sensing:** monitor the status and locations of troops, weapons, and supplies. Detect, locate or track enemy movements. Increase alertness to potential terrorist threats. Monitor and control civilian populations. Provide security in a shopping mall, parking garage or at some other facility.

- **Asset Tracking and Supply Chain Management:** tracking of goods and assets.

- **Intelligent Agriculture and Environmental Monitoring**: crop and livestock management, habitat monitoring and disaster detection.

- **Health Monitoring and Surgery:** physiological data such as body temperature, blood pressure, and pulse are sensed and automatically transmitted to a computer or physician.

- **Civil Engineering**: detect and warn of structural weakness (bridges), track groundwater patterns and how much carbon dioxide cities are expelling. Monitor traffic conditions and plan routes effectively. Determine which spots are occupied and which spots are free in car parks, etc.

The rest of the paper is organised as follows: Section 2 discusses security in WSNs and section 3 introduces the notion of trust. Trust and reputation models in different domains are presented in section 4 and section 5 concludes the paper.

## 2. SECURITY IN WSNs

In general, the key security goals of any network are to protect the network against all sorts of attacks, such as eavesdropping, fabrication, injection and modification of packets, impersonation; node capturing and many others, and to address other issues, like privacy, availability, accountability, data authentication, data integrity and data freshness. All these issues apply to traditional and wireless networks, but can have different consequences in WSNs, due to the open transmission medium, the resource constraints and the mass unattended deployment, especially in difficult and dangerous environments.

Research continues to be conducted into the design and optimisation of WSNs, as the use of these networks is still in its infancy phase. The security issue has been raised by many researchers [14-24], and, due to the deployment of WSN nodes in hazardous and/or hostile areas in large numbers, such deployment forces the nodes to be of low cost and therefore less reliable or more prone to overtaking by an adversary force. Some methods used, such as cryptographic authentication and other mechanisms [25-32], do not entirely solve the problem. For example, adversarial nodes can have access to valid cryptographic keys to access other nodes in the network. The reliability issue is certainly not addressed when sensor nodes are subject to system faults. These two sources of problems, system faults and erroneous data or bad routing by malicious nodes, can result in the total breakdown of a network and cryptography by itself is insufficient to solve these problems. So new tools from different domains─social sciences, statistics, e-commerce and others ─ should be integrated with cryptography to completely solve the unique security attacks in WSNs, such as node capturing, Sybil attacks, denial of service attacks, etc.

To protect the network from the above-mentioned attacks, a secure routing protocol (SRP), which addresses the limitation of sensor networks, must be used to secure the communication channel between nodes [33]; since routing in WSNs is a cooperative process whereby route information is relayed between nodes. As there is no guarantee that all nodes in the discovered route will behave as expected to fulfil the promises made, some malicious or selfish nodes might exist. That will lead to a network malfunction or breakdown and will require the SRP to discover and isolate the problem nodes, as the survivability of a WSN requires robustness against rapidly changing topologies and malicious attacks. There are different approaches followed by researchers [21] targeting MANETs and WSNs to solve this problem:

- Maintaining a trust and reputation table for all nodes in a sub-network
- The use of a watchdog mechanism to monitor the behaviour of all the surrounding nodes
- Discover faulty and/or misbehaving nodes, report them and exclude them from the network
- Provide incentives to nodes, so they comply with protocol rules
- The use of low-cost cryptography to protect the integrity and authenticity of exchanged data, without placing any overhead at intermediate nodes

All these approaches will be discussed in detail later in this paper.

### 3. NOTION OF TRUST

Due to the nature of WSN deployment being prone to the surrounding environment and suffering from other types of attacks in addition to the attacks found in traditional networks, other security measurements different from the traditional approaches must be in place to improve the security of the network. The trust establishment between nodes is a must to evaluate the trustworthiness of other nodes, as the survival of a WSN is dependent upon the cooperative and trusting nature of its nodes.

Security and trust are two tightly interdependent concepts and because of this interdependence, these terms are used interchangeably when defining a secure system [34]. However, security is different from trust and the key difference is that, it is more complex and the overhead is high.

Trust has been the focus of researchers for a long time [35], from the social sciences, where trust between humans has been studied to the effects of trust in economic transactions [36-38]. Although intuitively easy to comprehend, the notion of trust has not been formally defined. Unlike, for example, reliability, which was originally a measure of how long a machine can be trustworthy, and came to be rigorously defined as a probability, trust is yet to adopt a formal definition.

Along with the notion of trust, comes that of reputation [39], which is occasionally treated by some authors as trust. Reputation is not to be confused with trust: the former only partially affects the latter. Reputation is the opinion of one person about the other, of one internet buyer about an internet seller, and by construct, of one sensor node about another. Trust is a derivation of the reputation of an entity. Based on the reputation, a level of trust is bestowed upon an entity. The reputation itself has been built over time based on that entity's history of behaviour, and may be reflecting a positive or negative assessment. It is these quantities that researchers try to model and apply to security problems in WSNs.

Among the motivating fields for the development of trust models is e-commerce, which necessitated the notion of judging how trusted an internet seller can be [39, 40]. This was the case for peer-to-peer networks and other internet forums where users deal with each other in a decentralised fashion [41-45]. Recently, attention has been given to the concept of trust to increase security and reliability in ad-hoc [34, 46-53] and sensor networks [54-56]. WSNs are open to unique problems, due to their deployment and application nature. The low cost of the sensor nodes of a WSN prohibits sophisticated measures to ensure data authentication. Some methods used, such as cryptographic, authentication and other mechanisms [25-29, 31], do not entirely solve the problem. In the following section a brief survey, introducing only the methodology used to formulate trust and how is it being updated, of existing research on trust from different disciplines is presented in order to easily understand the concept of trust. More attention will be given to research being conducted on trust in MANETs and WSNs, as this is the area of interest for this research.

## 4. TRUST AND REPUTATION IN DIFFERENT DOMAINS

Understanding the notion of trust is the key to model trust properly in a specific discipline. Trust is an old but important issue in any networked environment; it has been there all of time and people have been using it in their daily life interactions without noticing, that is, buying and selling, communicating, cooperating, decision-making etc. involves some sort of trust without paying attention to it as a specific phenomenon.

### 4.1. Trust in Social Science and E-Commerce

Social science is concerned with the relationships between individuals in a society [57]. The concept of reputation in social networks is a natural one and people experience it in everyday life (buying, selling). Trust in general simplifies daily life by helping to solve some complex processes and by enabling the delegation of tasks and decisions to other trusted parties [58].

Reputation and trust systems in the context of e-commerce systems, such as eBay [39], Yahoo auctions [39], and internet-based systems such as Keynote [42, 44], use a centralised trust authority to maintain the reputation and trust values. Additionally, these systems use several debatable heuristics for the key steps of reputation updates and integration; due to the use of deterministic numbers for representing reputation [56].
Abdul-Rahman and Hailes in [59] proposed a model for supporting trust in virtual communities, based on direct experiences and reputation. Their model is based on a word-of-mouth mechanism, which allows agents to decide which other agents' opinion they trust more. They use direct and indirect (recommendations) trust and they introduced the semantic distance of the ratings in their mode.

In [60] Josang and Ismail developed the beta reputation system for electronic markets, based on distribution by modelling reputation as posterior probability based on past experiences. They used the beta probability density functions to combine feedback and derive reputation ratings. The advantages of the beta reputation system are flexibility and simplicity, as well as its foundation on the theory of statistics. The certainty of the trust calculation is defined by mapping the beta distribution to an opinion, which describes beliefs about the truth of statements.
Sabater and Sierra proposed the reputation system ReGreT in [61], which uses direct experiences, witness reputation and analysis of the social network in which the subject is embedded, to calculate trust. Trust is calculated as a weighted average of these experiences and uses the same value range. In [62] and [63], trust is explicitly based on rating experiences and integrating the rating into the existing trust. In [64] and [65], the authors indicated that "trust is more than a subjective probability" and presented a model of trust based on fuzzy logic that took into account many different beliefs and personality factors. On the topic of reputation, the authors of [66] presented a model in which agents can revise their beliefs based on information gathered from other agents and use Bayesian belief networks to filter out information they regard as false.

### 4.2. Trust in Distributed and Peer-to-Peer Systems

Reputation and trust systems in the context of distributed and peer-to-peer (P2P) networks are distributed; there is no centralised entity to oversee the behaviour of nodes in a network, so users keep track of their peers' behaviour and exchange this information directly with others; and also maintain a statistical representation of the reputation by borrowing tools from the realms of game theory [43], Bayesian networks [67-69] and other domains. These systems try to counter selfish routing misbehaviour of nodes by enforcing nodes to cooperate with each other.

Aberer and Despotovic in [41] were one of the first researchers to propose a reputation-based management system for P2P systems. However, their trust metric simply summarises the complaints a peer receives and it is very sensitive to the misbehaviour of peers. The resurrecting duckling model in [15] and its descendants [70, 71] represent a peer-to-peer trust establishment framework in which principals authenticate their communication channel by first exchanging keying material via an out-of-band physical contact. The established trust is binary; the communication channel is either secure or not secure and the model is based upon a hierarchical graph of master-slave relationships and is most suitable for security in large-scale dumb sensor nodes where pre-configuration has to be avoided.

The SECURE project [72, 73] (Secure Environments for Collaboration among Ubiquitous Roaming Entities) attempts to combine all aspects of trust-modelling into a single framework, ranging from trust-modelling and risk analysis to entity recognition and collaboration models [74]. The SECURE trust model extends the work of Weeks [75] in formalising trust management in security access control systems in terms of least fixed-point calculations, into evidence-based trust models. The model proposed allows each principal to express it's trust policy as a mathematical function which determines it's trust in everyone else, in terms of everyone else's trust assignments. These trust policies can then be combined to produce a consistent trust assignment for all participating principals.

The Distributed Trust Model proposed in [76] makes use of a protocol to exchange, revoke and refresh recommendations about other entities. By using a recommendation protocol, each entity maintains its own trust database. The proposed model is suitable for less formal, provisional and temporary trust relationships and adopts an averaging mechanism to yield a single recommendation value. The handling of false or malicious recommendations has to be supported via some out-of-band mechanism.

Wang and Vassileva in [67-69] modelled trust using Bayesian networks based on the quality of services provided by agents. An agent broadly builds two kinds of trust in another agent; one is the trust in another agent's competence to provide services and the other is the trust in another agent's reliability in providing recommendations about other agents. The systems use binary events — successful or unsuccessful transactions to model trust and weight the direct and indirect information differently. According to the authors, the node will discard the recommendations from the untrustworthy sources but will combine the recommendations from the trustworthy and unknown sources. Here the reliability includes two aspects: whether the agent is truthful in telling its information and whether the agent is trustworthy or not. Although the models presented in [77] and [78] are based on the quality of services provided by the other peer, they model trust as a weighted vector of all services provided and each service is weighted differently based on its importance.

Kinateder et al., in [79] presented a good comparison between some trust update algorithms and proposed a generic trust model "UniTEC", which provides a common trust representation for trust update algorithms based on experiences. They used direct and indirect trust update separately, by giving more weight to the recent experience than to the old one. This calculates a new trust value based on the old trust value and the new rating. Ratings in the original UniTEC proposal are expressed as a binary metric of either bad or good experience. Azzedin and Maheswaran's trust model in [80] computed the eventual trust based on a combination of direct trust and reputation, by weighting the two components differently; giving more weight to the direct trust. The trust level is built on past experiences and is given for a specific context. Other authors, such as [81] and [82], proposed a neural network approach to model reputation in distributed systems.

BambooTrust, presented in [83], is a practical and high-performance distributed trust management system for global public computing platforms such as grid computing systems. It is based on the XenoTrust presented in [84] and the Bamboo hash table and is implemented to facilitate performance, scalability, efficiency and load-balancing. BambooTrust is built as a peer-to-peer system in which nodes are identical in terms of what they do, and it also requires a public and a private key to be used. XenoTrust [84] uses the criteria of performance (reliability, honesty and throughput) to assess the others. It is an event-based distributed trust management used in the Xenoserver open platform. Most of the existing trust management systems depend on the traditional request/reply paradigm, which involves polling and causes communication overhead, while the event-based depends on whether a change has occurred or not.

Shand et al., in [85], presented a trust framework to facilitate secure collaboration in pervasive computer systems. Most trust models before this model used security policy, which permits or prohibits the actions and these policy-based models are not suitable for dynamic networks, with topology changing all the time. For each entity recommendation a policy function is formed and by combining these policy functions with other policy functions, the trust level is calculated. Daniele et al., who presented the B-trust model in [86], proposed a lightweight distributed trust framework for pervasive computing that evolves trust based on a Bayesian formalisation, and protects user anonymity, whilst being resistant to "Sybil attacks". They used the weighting approach to weight the direct and indirect (both good and bad) information.

### 4.3. Trust in Ad-hoc Networks

Ad-hoc networks are characterised by dynamically changing their structure; this means nodes join and leave networks very often. While in a roaming process nodes are continuously confronted with other (unknown) nodes, which can be of a great help to them if they can collaborate with each other, collaboration between strange nodes is not fully utilised, due to the fear of not being trusted and the potential risk of such collaboration. Trust relationships in MANETs are established, evolved, propagated and expired on the fly (no infrastructure) and are very susceptible to attacks, as the whole environment is vulnerable due to the shared wireless medium. In other words, there is no a priori trusted subset of nodes to support the network functionality. Trust may only be developed over time, while trust relationships among nodes may also change [21].

Reputation and trust systems in the context of ad-hoc networks, CONFIDANT [50] and CORE [51], maintain a statistical representation of the reputation by borrowing tools from the realms of Bayesian estimation and game theory respectively. These systems try to counter selfish routing misbehaviour of nodes by enforcing nodes to cooperate with each other. Recently proposed reputation systems in the domain of ad-hoc networks formulate the problem in the realm of Bayesian analytics rather than game theory [87, 88].

Michiardi and Molva, in [51], proposed the CORE system (A Collaborative Reputation Mechanism to enforce node cooperation in MANETs), which uses game theoretic analysis to model reputation. Members that have a good reputation, because they helpfully contribute to the community life, can use the resources, while members with a bad reputation, because they refuse to cooperate, are gradually excluded from the community. In CORE [51], the term "subjective reputation" is used to represent the reputation calculated from direct observations' using a weighted mean of the observations rating factors, giving more relevance to the past observations. The system uses only the positive value of the indirect reputation to prevent the badmouthing attacks, but does not address the issue of collusion to create false praise. The term "functional reputation" is introduced to allow the subjective and the indirect reputation to be calculated with regard to different functions and to combine them using different weights to obtain a global reputation value.

The CONFIDANT Protocol (Cooperation of Nodes, Fairness In Dynamic Ad-hoc Networks) proposed in [50] is based on direct observations and on second-hand information from other nodes and is updated according to a Bayesian estimation. The model shows that using second-hand information can significantly accelerate the detection and subsequent isolation of malicious nodes. The authors later improved CONFIDANT with an adaptive Bayesian reputation and trust system in [89] and [90]. CONFIDANT differs from CORE only in that it sends reputation values to other nodes in the network, which exposes the scheme to malicious spreading of false reputation values. If a node is observed to behave in a cooperative fashion, then positive reputation is assigned to it, otherwise a negative reputation is assigned to it.

Theodorakopoulos and Baras' model, in [91], presented an extension of their previous work in [92], using the theory of semirings to evaluate the process, which is modelled as a path problem on a directed graph, where nodes represent entities and edges represent trust relations. In their model, users can use the second-hand information from the intermediate nodes to form their opinion about the others, which means they do not need to have a direct connection with the other nodes to deal with them. The second-hand information is not as valuable as the direct interactions. Also Baras et al., in [47], proposed a solution to the problem of establishing and maintaining trust in MANETs, which addresses the dynamism and the resource constraints of such a network. The network in [47] is modelled as an undirected graph (nodes and links), which is based on the Ising model in physics [93]; nodes interact only with their neighbours.

Buchegger and Boudec, in [89], proposed a system which is robust to false ratings: accusation or praise. Every node maintains a reputation rating and trust rating for other one, and the first-hand information is exchanged with others from time to time. Using a modified Bayesian approach, only the second-hand reputation information that is compatible with the current reputation information is accepted. Reputation rating is modified based on the accepted information and trust rating is updated based on reputation rating. They introduced the re-evaluation process and reputation fading to prevent the exploitation of good reputation built over time. In their model, only the first-hand information is published and, to allow for reputation fading, they weight the evidence by time, giving less weight to observations received in the past. This is different from the standard Bayesian, which gives the same weight for all observations regardless of their time of occurrence; that is why the system is a modified Bayesian approach. To accelerate the detection of misbehaving nodes, the authors used selected second-hand information, which comes from continuously trusted nodes, or passed the deviation test which evaluates compatibility with its own reputation ratings.

Pirzada and McDonald introduced the notion of belief in their communication trust-based model in [34], which provides a dynamic measure of reliability and trustworthiness suitable for applications in an ad-hoc network. The trust model in [34] is an adaptation of Marsh's [94] trust model, but they merged utility and importance in one variable called weight for simplicity. They categorised trust into different categories and calculated trust as a sum of all these weighted categories. Based on the protocol and on the scenario to which the trust model is applied, the total number of categories was defined. The main goal of their model was to build route trustworthiness in nodes by extending the dynamic source protocol to receive a complete list of all nodes through which a protocol has passed.

Trust as a measure of uncertainty was presented in [95] by Sun et al., and as such trust values can be measured by entropy. From this understanding of trust, a few techniques were developed to calculate trust values from observation. In addition, two models were designed: the entropy-based trust model and probability-based trust model to address the concatenation and multi-path trust propagation problems in ad-hoc networks. The proposed models investigate trust relationship associated with packet forwarding, as well as making recommendations.

Liu et al., model, proposed in [33], is used to determine and maintain dynamic trust relationships and then make routing decisions. The purpose of the model is to enhance the security of message routing in MANETs, via selecting the most secure route based on the determined and maintained trust values between nodes. It is based on the assumptions that every node deployed possesses an intrusion detection system (IDS) that can detect and report the behaviour of malicious nodes, and that nodes are stimulated to cooperate adequately on the network. Each node in the network is initially authenticated by an authentication mechanism and is assigned with a trust value, which is updated automatically based on reports from the IDS.

Davis, in [49], presented a trust management scheme based on a structured hierarchical trust model, which addresses the explicit revocation of certificates and, as they claimed their scheme is robust against malicious accusation exploits (one node accuses the other as malicious, whereas it is not malicious). The model mainly uses digital certificates to establish trust and for a node to be trusted, it needs to present an active certificate which has never been revoked, which means every node is pre-deployed with a certificate. The model is based on assigning different weights to different accusations, and if the sum of all weighted accusations is greater than a pre-defined threshold, the certificate should be revoked.

Models in [96] and [97] propose probabilistic solutions based on a distributed trust model to establish trust relationships between nodes in an ad-hoc network, which does not rely on any previous assumptions. In [96], the authors used the idea of directed graph as their mathematical representations, while in [97], they are using the Beta distribution to calculate trust in MANETs. Jiang and Baras' trust model, presented in [98], consists of two components: a trust computational model (evaluation model) and a trust evidence distribution model (the input of the evaluation model). The model uses the swarm intelligence paradigm and ideas from a P2P file-sharing system. The model mainly addresses the evidence distribution and retrieval system, using the public and private key concepts, and uses certificates to distribute the evidence.

### 4.4. Trust in Sensor Networks

Trust in WSN networks plays an important role in constructing the network and making the addition and/or deletion of sensor nodes from a network, due to the growth of the network, or the replacement of failing and unreliable nodes very smooth and transparent. The creation, operation, management and survival of a WSN are dependent upon the cooperative and trusting nature of its nodes, therefore the trust establishment between nodes is a must. However, using the traditional tools such as cryptographic tools to generate trust evidence and establish trust and traditional protocols to exchange and distribute keys is not possible in a WSN, due to the resource limitations of sensor nodes [46]. Therefore, new innovative methods to secure communication and distribution of trust values between nodes are needed. Trust in WSNs, has been studied lightly by current researchers and is still an open and challenging field.

Reputation and trust systems in the context of sensor networks prior to this research have received little attention from researchers, however, recently researchers have started to make efforts on the trust topic, as sensor networks are becoming more popular. Ganeriwal and Srivastava were the first to introduce a reputation model specific to sensor networks in [54]; the RFSN (Reputation-based Framework for High Integrity Sensor Networks) model uses the Beta distribution, as a mathematical tool to represent and continuously update trust and reputation. The model classifies the actions as cooperative and non-cooperative (binary) and uses direct and indirect (second-hand) information to calculate the reputation. The second-hand information is weighted by giving more weight to the information coming from very reliable nodes. Trust is calculated as an expected value of the reputation and the behaviour of the node is decided upon a global threshold; if the trust value is below a threshold, the node is uncooperative, otherwise it is cooperative. The system propagates only the positive reputation information about other nodes [54], and by doing so, it eliminates the bad-mouthing attack, but at the same time it will affect the system's efficiency, as nodes will not be able to exchange their bad experience with malicious nodes. The aging factor is also introduced to differently weight the old and new interactions; more weight is given to recent interactions.

The DRBTS (Distributed Reputation-based Beacon Trust System) presented in [55] is an extension to the system introduced in [99], which presented a suite of techniques that detect and revoke malicious beacon nodes that provide misleading location information. It is a distributed security protocol designed to provide a method in which beacon nodes can monitor each other and provide information so that sensor nodes can choose to trust, using a voting approach. Every beacon node monitors its one hope neighbourhood for misbehaving beacon nodes and accordingly updates the reputation of the corresponding beacon node in the neighbour-reputation table. Beacon nodes use second-hand information for updating the reputation of their neighbours after the second-hand information passes a deviation test. A sensor node uses the neighbour-reputation table to determine whether or not to use a given beacon's location information based on a simple majority voting scheme. The DRBTS models the network as an undirected graph, uses first-hand and second-hand information to build trust.

Garth et al., [100] proposed a distributed trust-based framework and a mechanism for the election of trustworthy cluster heads in a cluster-based WSN. The model uses direct and indirect information coming from trusted nodes. Trust is modelled using the traditional weighting mechanism of the parameters: packet drop rate, data packets and control packets. Each node stores a trust table for all the surrounding nodes and these values are reported to the cluster head only and upon request. This approach is not based on second-hand information, so it reduces the effect of bad-mouthing. Hur et al., proposed a trust model in [101], to identify the trustworthiness of sensor nodes and to filter out (remove) the data received from malicious nodes. In their model, they assume that each sensor node has knowledge of its own location, time is synchronised and nodes are densely deployed. They computed trust in a traditional way, weighting the trust factors (depending on the application) and there is no update of trust.

The proposed reputation-based trust model in WSNs by Chen et al., in [102], borrows tools from probability, statistics and mathematical analysis. They argued that the positive and/or negative outcomes for a certain event are not enough to make a decision in a WSN. They built up a reputation space and trust space in WSNs, and defined a transformation from the reputation space to the trust space [102]. The same approach presented in RFSN [54] is followed; a watchdog mechanism to monitor the other nodes and to calculate the reputation and eventually to calculate trust, and Bayes' theorem is used to describe the binary events, successful and unsuccessful transactions, with the introduction of uncertainty. Initially, the trust between strangers is set to (0) and the uncertainty is set to (1). The model does not use second-hand information, and how to refresh the reputation value is an issue. Xiao et al., in [103] developed a mechanism called SensorRank for rating sensors in terms of correlation by exploring Markov Chains in the network. A network voting algorithm called TrustVoting was also proposed to determine faulty sensor readings. The TrustVoting algorithm consists of two phases: self diagnose (direct reading) and neighbour diagnose (indirect reading), and if the reading is faulty then the node will not participate in the voting.

Crosby and Pissinou, in [104], proposed a secure cluster formation algorithm to facilitate the establishment of trusted clusters via pre-distributed keys and to prevent the election of compromised or malicious nodes as cluster heads. They used Beta distribution to model trust, based on successful and unsuccessful interactions. The updating occurs through incorporating the successful/unsuccessful interactions at time *t+1* with those of time *t*. Their trust framework is designed in the context of a cluster-based network model with nodes that have unique local IDs. The authors of [105] proposed the TIBFIT protocol to diagnose and mask arbitrary node failures in an event-driven wireless sensor network. The TIBFIT protocol is designed to determine whether an event has occurred or not through analysing the binary reports from the event neighbours. The protocol outperforms the standard voting scheme for event detection.

A few other systems related to trust in WSNs, have been proposed in the literature such as [106-112], which use one or more of the techniques mentioned before to calculate trust. The proposed model in [106] uses a single trust value for a whole group (cluster), assuming that sensor nodes mostly fulfil their responsibilities in a cooperative manner rather than individually. In [107], the model is based on a distributed trust model to produce a trust relationship for sensor networks and uses the weighting approach to combine trust from different sources. In [108], a trust-based routing scheme is presented, which finds a forwarding path based on packet trust requirements, also using the weighting approach. In [109], a stochastic process formulation based on a number of assumptions is proposed to investigate the impact of liars on their peers' reputation about a subject. In [110], the authors proposed a new fault-intrusion tolerant routing mechanism called MVMP (multi-version multi-path) for WSNs to provide both fault tolerance and intrusion tolerance at the same time. The proposed model in [111] is an application-independent framework, built on the alert-based detection mechanisms provided by applications, to identify the malicious (compromised) nodes in WSNs. In [112], a parameterised and localised trust management scheme for sensor networks security (PLUS) is presented, whereby each sensor node rates the trustworthiness of its interested neighbours, identifies the malicious nodes and shares the opinion locally.

From the above survey, which has introduced most of the work undertaken in the area of trust in different domains, it can be noticed that researchers are using many types of methodologies borrowed from different domains to calculate trust, based on different criteria. Table 1 below summarises most of the above-mentioned trust models and the methodologies used to model trust, also the factors used to update trust, direct and indirect trust are summarised in Table 2.

Table 1. Methodologies used to model trust and their references

| Methodology | References |
|---|---|
| Ratings | [39, 59, 62, 63, 89] |
| Weighting | [34, 61, 77, 78, 80, 94, 100, 107, 108] |
| Probability | [50, 54, 60, 87-89, 95-97, 102-104, 109] |
| Bayesian network | [66-69, 86] |
| Neural network | [81, 82] |
| Game theory | [43, 51] |
| Fuzzy logic | [64, 65] |
| Swarm intelligence | [15, 70, 71, 98] |
| Directed and undirected graph | [47, 55, 91-93] |

Table 2. Factors used to updating trust and their references

| Factor | References |
|---|---|
| Direct only | [102] |
| Indirect only | [89, 100] |
| Indirect positive | [51, 54] |
| Both | [50, 55, 59-61, 67-69, 80, 86] |

It is also worth mentioning that almost all the work undertaken on trust is based on successful and unsuccessful (binary) transactions between entities, that is, trust has been modelled in networks in general from a communication point of view, with no exception for WSNs, which is characterised by a unique feature: sensing events and reporting data. This unique characteristic is the basis of our research, which is focusing on modelling and calculating trust between nodes in WSNs based on continuous data (sensed events) and will eventually introduce the communication as a second factor of trust. Accordingly, a trust classification for WSNs has been introduced in [113] and in [114, 115] a new framework to calculate trust in WSNs has been introduced, using the traditional weighting approach to combine direct and indirect trust. In [116], the sensed data was introduced as the decisive factor of trust, that is, trust in WSNs was modelled from the sensor reliability perspective. The RBATMWSN model introduced in [117], represents a new trust model and a reputation system for WSNs, based on sensed continuous data. The trust model establishes the continuous version of the Beta reputation system applied to binary events and presents a new Gaussian Trust and Reputation System for Sensor Networks (GTRSSN), as introduced in [118], which introduces a theoretically sound Bayesian probabilistic approach for mixing second-hand information from neighbouring nodes with directly observed information to calculate trust between nodes in WSNs, and finally a Bayesian fusion approach was introduced in [119], to combine continuous data trust based on sensed events and binary communication trust based on successful and unsuccessful transactions between nodes.

## 5. CONCLUSION

Trust as an essential attribute in building a relationship between entities has been studied for a long time by scientists from disparate scientific fields. Every field has examined modelling and calculating trust using different techniques, and one of the most prominent and promising techniques is the use of statistics, mainly probabilities to solve the problem, especially in dynamic networks where the topology is changing rapidly.

This paper has briefly introduced wireless sensor networks and the challenges associated with deploying them in unattended and difficult environments. It has also introduced the security issues in WSNs and the need for new innovative approaches to solve these issues. In the notion of trust, the difference between trust and security has been discussed and it has been explained that trust is not the same as security, even though they are sometimes used interchangeably to describe a secure system. The difference between reputation and trust has also been discussed; the former only partially affects the latter, which means, based on reputation, a level of trust is bestowed upon an entity.

A concise and closely related survey of trust models in different domains — social science and e-commerce, distributed and peer-to-peer networks, ad-hoc networks and wireless sensor networks has also been presented, showing the methodology used to formulate trust in each model, and the way in which the trust updating process is achieved has also been discussed and summarised. Finally, the survey has also shown that, even though researchers have started to explore the issue of trust in wireless sensor networks, they are still following almost the same approaches used by researchers in other fields to model trust; examining the issue of trust from a binary communication point of view (routing). This is in contrast to our research, which takes into consideration not only the communication side but also the continuous sensed data, which is a unique characteristic of sensor networks and has never been addressed by trust researchers in WSNs.

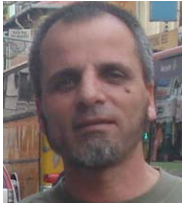

Mohammad Momani is a sessional lecturer at the University of Technology, Sydney, Faculty of Engineering and Information Technologies. His research interests are trust management and security issues in mobile ad hoc networks and wireless sensor networks. He graduated with a PhD in Computer Engineering from the University of Technology, Sydney, 2009, a M.Sc. in Internetworking from the University of Technology, Sydney, 2003 and a M.Sc. in Computer Engineering from Bulgaria, 1986. He is also a CCNI, CCNA, MCSE and a CNE certified with almost 20 years of experience in the computer industry.

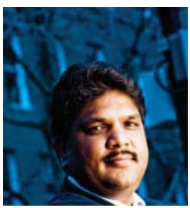

Prof. Subhash Challa is the Senior Principal Scientist at NICTA Victorian Research Lab, Melbourne, Australia. Prior to this he was the Professor of Computer Systems at the University of Technology, Sydney, and leads the Networked Sensor Technologies (NeST) Lab. He was a senior research fellow at the University of Melbourne where he led a number of tracking and data fusion projects in collaboration with DSTO, DARPA, BAE, Raytheon, Tenix Defense, Thales, RTA, NSW Police and others. He received his PhD from QUT, Brisbane, Australia in 1999. Part of his PhD was completed at Harvard Robotics Lab, Harvard University, Boston, USA. Subhash was a Tan-Chun-Tau Fellow at Nanyang Technological University, Singapore 2002-2003. He has been the plenary, tutorial and invited speaker at various information fusion and sensor network conferences worldwide, including IDC 2007 (Adelaide, Australia), ensors Expo 2006 (Chicago, USA), ISSNIP conferences (2005, 2004) and Fusion Conferences (2003, 2005, 2006). He is co-authoring of reference text "Fundamentals of Object Tracking," to be published by Cambridge University Press, UK. He is also the associate editor of the Journal of Advances in Information fusion. He has published about 70 papers in various International journals and conferences.